\documentclass[a4paper]{jpconf}
\usepackage{amsfonts,amssymb,amsmath,amsthm,cite}
\usepackage{graphicx}
\DeclareMathOperator{\Ker}{Ker}          
\DeclareMathOperator{\Res}{Res}          

\newtheorem{assumption}{Assumption}[section]
\newtheorem{theorem}[assumption]{Theorem}

\newtheorem{conjecture}[assumption]{Conjecture}
\newtheorem{lemma}[assumption]{Lemma}
\newtheorem{definition}[assumption]{Definition}
\newtheorem{prop}[assumption]{Proposition}
\newtheorem{remark}[assumption]{Remark}

\newcommand{\A}{\mathcal{A}}             
\newcommand{\B}{\mathcal{B}}             
\newcommand{\C}{\mathbb{C}}              
\newcommand{\Coo}{C^\infty}              
\newcommand{\DD}{\mathcal{D}}            
\newcommand{\eps}{\varepsilon}           
\newcommand{\ga}{\gamma}                 
\renewcommand{\H}{\mathcal{H}}           
\newcommand{\half}{{\mathchoice{\thalf}{\thalf}{\shalf}{\shalf}}}
\renewcommand{\L}{\mathcal{L}}           
\newcommand{\N}{\mathbb{N}}              
\newcommand{\pa}{\partial}
\newcommand{\Q}{\mathbb{Q}}              
\newcommand{\R}{\mathbb{R}}              

\newcommand{\set}[1]{\{\,#1\,\}}         
\newcommand{\shalf}{{\scriptstyle\frac{1}{2}}}  
\renewcommand{\SS}{\mathcal{S}}          
\newcommand{\T}{\mathbb{T}}              
\newcommand{\Th}{\Theta}
\renewcommand{\th}{\theta}
\newcommand{\thalf}{\tfrac{1}{2}}        
\newcommand{\wt}{\widetilde}             
\newcommand{\Z}{\mathbb{Z}}              

\def\<#1,#2>{\langle#1\,,\,#2\rangle}        

\newbox\ncintdbox \newbox\ncinttbox
\setbox0=\hbox{$-$} \setbox2=\hbox{$\displaystyle\int$}
\setbox\ncintdbox=\hbox{\rlap{\hbox
    to \wd2{\kern-.1em\box2\relax\hfil}}\box0\kern.1em}
\setbox0=\hbox{$\vcenter{\hrule width 4pt}$}
\setbox2=\hbox{$\textstyle\int$}
\setbox\ncinttbox=\hbox{\rlap{\hbox
    to \wd2{\kern-.14em\box2\relax\hfil}}\box0\kern.1em}
\newcommand{\ncint}{\mathop{\mathchoice{\copy\ncintdbox}
    {\copy\ncinttbox}{\copy\ncinttbox}
    {\copy\ncinttbox}}\nolimits}

\begin{document}

\title{Spectral action in noncommutative geometry: An example} 

\author{B. Iochum$^{1,2}$}

\address{$^1$Centre de Physique Th\'eorique
\footnote[3]{Unit\'e Mixte de Recherche (UMR 6207) du CNRS et des
Universit\'es Aix-Marseille I, Aix-Marseille II et de l'Universit\'e
du Sud Toulon-Var. Laboratoire affili\'e \`a la FRUMAM -- FR 229

CPT-P27-2007}, 
CNRS--Luminy, Case 907,13288 Marseille Cedex 9 FRANCE}

\address{$^2$ Also at Universit\'e de Provence}

\ead{iochum@cpt.univ-mrs.fr}


\begin{abstract}
This is a report on a joint work \cite{EILS} with D. Essouabri, C. Levy and A. Sitarz. 

The spectral action on noncommutative torus is obtained, using a
Chamseddine--Connes formula via computations of zeta functions.
The importance of a Diophantine condition is outlined as far as the difficulties to go beyond. Some results on holomorphic continuation of series of holomorphic functions are presented.
\end{abstract}

\section{Introduction}

The spectral action introduced by
Chamseddine--Connes plays an important role (see \cite{CC} and Chamseddine's contribution to this proceedings) in
noncommutative geometry. More
precisely, given a spectral triple $(\A,\H,\DD)$ where $\A$ is an
algebra acting on the Hilbert space $\H$ and $\DD$ is a Dirac-like
operator (see \cite{Book,Polaris}), they proposed a physical action
depending only on the spectrum of the covariant Dirac operator
\begin{equation}
\label{covDirac} 
\DD_{A}:=\DD + \wt{A}, \quad \wt{A}:=A + \epsilon \,JAJ^{-1}
\end{equation}
where $A$ is a one-form represented on $\H$,
so has the decomposition $A=\sum_{i}a_{i}[\DD,\,b_{i}]$, 
with $a_{i}$, $b_{i}\in \A$, $J$ is a real
structure on the triple corresponding to charge conjugation and
$\epsilon \in \set{1,-1}$ depending on the dimension of this triple
and comes from the commutation relation
\begin{equation}
    \label{Jcom}
J\DD=\epsilon \, \DD J.
\end{equation}

This action is defined by
\begin{equation}
\label{action}
\SS(\DD_{A},\Phi,\Lambda):=\Tr \big( \Phi( \DD_{A} /\Lambda) \big)
\end{equation}
where $\Phi$ is any even positive cut-off function which could
be think as a step function. This means that $\Phi$ counts the
spectral values of $\vert \DD_{A} \vert$ less than the mass scale
$\Lambda$.

Even if the spectral action on NC-tori has been computed
for operators of the form $\DD + A$ in  \cite{GI2002} and for $\DD_{A}$
in \cite{GIVas}, it is interesting to show that this can be also directly obtained from the Chamseddine--Connes analysis of \cite{CC1}. Actually,
\begin{align}
\label{formuleaction}
    \SS(\DD_{A},\Phi,\Lambda) \, = \,\sum_{0<k\in Sd^+} \Phi_{k}\,
    \Lambda^{k} \ncint \vert D_{A}\vert^{-k} + \Phi(0) \,
    \zeta_{D_{A}}(0) +\mathcal{O}(\Lambda^{-1})
\end{align}
where $D_A = \DD_A + P_A$, $P_A$ is the projection on $\Ker \DD_A$,
$\Phi_{k}= \half\int_{0}^{\infty} \Phi(t) \, t^{k/2-1} \, dt$ and
$Sd^+$ is the strictly positive part of the dimension
spectrum of $(\A,\H,\DD)$. As we will see,
$Sd^+=\set{1,2,\cdots,n}$ and $\ncint |D_{A}|^{-n}=\ncint |D|^{-n}$. Moreover, 
the constant term $\zeta_{D_A}(0)$ in (\ref{formuleaction}) can
be computed from unperturbed spectral action since it has been proved in \cite{CC1}  that
\begin{align}
\label{constant}
\zeta_{\DD+A}(0)-\zeta_{\DD}(0)= \sum_{q=1}^{n}\tfrac{(-1)^{q}}{q}
\ncint (A\DD^{-1})^{q},
\end{align}
using $\zeta_X(s)=\Tr(|X|^{-s})$. We show that this formula can be extended to non invertible Dirac operators and non invertible perturbations.

All results on spectral action are quite important in physics,
especially in quantum field theory and particle physics, where one
adds to the effective action some counterterms explicitly given by
(\ref{constant}), see for instance
\cite{Carminati,CC,CC1,CCM,Gayral,Goursac,GI2002,GIVas,
Knecht,Strelchenko,Vassilevich2002,Vassilevich2005,Vassilevich2007}.

\section{How to compute $\Tr \big( \Phi( \DD_{A} /\Lambda) \big)$ ?}
\subsection{Heat kernel approach}
\underline{Constraints on $\Phi$:} Actually, the step functions is not allowed due to Gibbs' phenomenon. So, we can 

- use the distributional approach investigated in \cite{Odysseus},

- or follow \cite{Nestetal} and assume that $\Phi \in C^{\infty}(\R^+)$  is a Laplace transform of 
$\hat \psi$ in the Schwartz space $\SS (\R^+)=\set{g \in \SS(\R)\, : \, g(x)=0, \, x \leq 0}$. 
So $\Phi$ has analytic extension on the right complex plane and 
$\Phi(z)=(-1)^m \int_0^{\infty} e^{-tz}\, t^m\; \hat \psi (t) \, dt$, $Re(z)>0$ where $ m:=\lfloor \tfrac n2 \rfloor$. 
In this case, the following is well established
\begin{align}
\label{heatexpansion}
\Tr \big( \Phi(\DD/\Lambda)\big) = \sum_{k=0}^{d} \Phi_k\,a_k\,\Lambda^{d-k} +\mathcal{O}(\Lambda^{-1})
\end{align}
where $a_k$ are the Seeley--De Witt coefficients (see \cite{newGilkey}) and 
when $n=2m$ is even, $\Phi_{2k}$ has the more familiar form:
\begin{eqnarray*}
\Phi_{2k}=
\begin{cases}
{\textstyle{\frac{1}{\Gamma(m-k)}}}\;\int_0^\infty \Phi(t) 
\; t^{m-1-k} \; dt,\;& {\rm for}\;
k=0,\cdots,m-1,\cr
(-1)^{k}\;\Phi^{(k-m)}(0),& {\rm for}\; k=m,\cdots,n.
\end{cases}
\end{eqnarray*}
For $n$ odd, the coefficients
$\Phi_{2k}$ have less explicit forms because 
they involve fractional derivatives of $\Phi$.

\subsection{Pseudodifferential operators and Zeta functions approach}
One can enter in the field either via the heat kernel as before or via zeta functions or Dixmier traces.

Consider first the commutative case where $\A \in C^{\infty}(M)$ for a manifold $M$. When $P \in \Psi DO$ is a pseudodifferential operator of order $q$
$$
\zeta_P(s)=\Tr (P\vert\DD\vert^{-s})
$$
is holomorphic for Re$(s)>q+d$, with at most poles at integers $k\leq q+d$.

The leading residue is
$$
\Res_{s=d+q} \, \zeta_P(s) = c_1 \Tr_{Dix} (P) = c_2 \int_{S^*(M)} \sigma_P
$$
(see Guillemin, Wodzicki,\ldots, where in general  $\DD^2=\triangle$ is the scalar Laplacian.)

Here is the interest of the abstract setting introduced in \cite{Cgeom,CM} where subleading residues are equally obtained: Given a spectral triple $(\A,\H,\DD)$ of dimension $d$, which means 
$$
0<\Tr_{Dix}(a\vert \DD \vert^{-d})<\infty \quad \forall  \, 0  \neq a \in \A^+
$$
(when $\DD$ is non invertible, one replace 
$\DD \rightarrow D:=\DD +P_0$ where $P_0$ is the projection on Ker  $\DD$,) one follow the construction of $\Psi DO$ on the track of Cordes' characterization of pseudodifferential operators:
\begin{align*}
OP^0&:=\set{T : t \rightarrow e^{it\vert \DD \vert} \, T \, e^{-it\vert \DD \vert} \in C^{\infty}\big(\R,\B(\H)\big)},\\
OP^{\alpha}&:=\set{T : T \vert D \vert^{-\alpha} \in OP^0}, \quad \alpha \in \R.
\end{align*}
\begin{definition}
\label{defpseudo} Define $\DD(\A)$ as the polynomial algebra generated
by $\A$, $J\A J^{-1}$, $\DD$ and $|\DD|$. A pseudodifferential operator is an operator $T$ such that there exists $d\in \Z$ such that
for any $N\in \N$, there exist $p\in \N_0$, $P\in \DD(\A)$ and $R\in OP^{-N}$ ($p$, $P$ and
$R$ may depend on $N$) such that $P\,D^{-2p}\in OP^d$ and
$$
T=P\,D^{-2p}+R\, .
$$
Define $\Psi(\A)$ as the set of pseudodifferential operators and
$\Psi(\A)^k:=\Psi(\A)\cap OP^k$.
\end{definition}
The idea is to work modulo large $N$.
This nonstandard definition pays attention to the reality operator $J$ and the kernel of $\DD$ and allows $\DD$ and $|D|^{-1}$
to be pseudodifferential operators. It is more in the spirit of \cite{CC1}.

Note that if $A$ is a 1-form, $A$ and $JAJ^{-1}$ are in $\DD(\A)$ and moreover
$\DD(\A)\subseteq \cup_{p \in \N_0} OP^p$. Since $|\DD|\in \DD(\A)$ by definition and $P_0$ is a pseudodifferential operator, for any
$p\in \Z$, $|D|^{p}$ is a pseudodifferential operator (in $OP^{p}$.) Remark also that $\DD(\A)\subseteq\Psi(\A) \subseteq \cup_{k\in \Z} OP^{k}$.

As in \cite{CM,Cgeom}, $\Psi(\A)$ is an algebra: if $T\in \Psi(\A)^d$, $T\in \Psi(\A)^{d'}$, then $TT'\in \Psi(\A)^{d+d'}$.

The noncommutative integral is naturally defined as 
$$
\ncint T := \Res_{s=0} \, \zeta_D^T(s) = \Res_{s=0} \Tr \big(T \, \vert D \vert^{-s} \big), \quad T \in \Psi DO.
$$
which makes sense since $a \vert D \vert^{-(d+\epsilon)} \in \L^1(\H)$, $\forall a \in \A$, $\forall \epsilon>0$.

One checks that $\ncint$ is a trace on $\Psi DO$ when all poles of $Sd(\A,\H,\DD)$ are simple.

Another important notion is the {\it dimension spectrum}
$$
Sd(\A,\H,D):=\set{\text{poles of } \Tr \big(T \vert D \vert^{-s}\big) : T \in \Psi DO^0}
$$
For example, when $M$ is a compact spin$^c$ Riemannian manifold of dimension $n$, with spectral triple 
$\A=C^\infty(M)$, $\H=L^2(Spineurs)$ and $\DD=-i\gamma^{\mu}\pa_{\mu}$, then $Sd(\A,\H,\DD)=\set{n-k : k \in \N_0}.$

\begin{remark}
It is rather difficult to compute zeta functions, even when it is a Dixmier-trace. However, it is proved  in \cite{CPRS} that  for any $0\leq T \in \B(\H)$ such that $T^s \in \L^1(\H)$ for all $s>1$, then, if $l=\lim_{\epsilon \rightarrow 0} \,\epsilon \Tr (T^{1+\epsilon})$ exists,  $T$ is Dixmier-traceable and $\Tr_{Dix} (T) =l$.
\end{remark}
\begin{remark}
When $\DD$ (or $\DD_A$) are not invertible, one uses in noncommutative integrals $D$ (or $D_A$) defined by $D=\DD+P_0$ (or $D_A=\DD_A+P_A$), where the $P$'s are projections on the kernel of the operator. Since $\DD$ has a compact resolvent,  $\DD_A$ has also a compact resolvent and these projections are finite-rank operators.

Note that $P_A \in OP^{-\infty}$, so is a smoothing operator.

For the kernels, there is a difference between $\DD_{A}$ and $\DD+A$: for the noncommutative torus, the inclusion $\Ker \DD \subseteq \Ker \DD +A$ is not satisfied since $A$ does not preserve $\Ker \DD$ contrarily to $\wt A$.
\end{remark}
Using (\ref{heatexpansion}), one obtains (\ref{formuleaction}). Moreover, relation \eqref{constant} proved in \cite{CC1},  can be extended to \begin{prop} 
For any selfadjoint one-form A, the constant term in $\Lambda$ in \eqref{heatexpansion} is
\begin{align}
\label{constant1}
\zeta_{D_A}(0)-\zeta_D(0)&=-\ncint \log(1+\wt A D^{-1})
=\sum_{q=1}^n \tfrac{(-1)^q}{q} \ncint (\wt A D^{-1})^q.
\end{align}
\end{prop}

\section{The noncommutative torus}
\subsection{Notations}

Let $\A_{\Th} :=\Coo(\T^n_\Th)$ be the smooth noncommutative
$n$-torus associated to a non-zero skew-symmetric deformation matrix
$\Th \in M_n(\R)$ (see \cite{ConnesTorus}, \cite{RieffelRot}): 
$\Coo(\T^n_\Th)$ is generated by $n$ unitaries $u_i$, $i=1,\dots,n$ subject to the relations $u_i\,u_j=e^{i\Th_{ij}}\,u_j\,u_i$, 
and with Schwartz coefficients: an element $a\in\A_{\Th}$ can be written as $a=\sum_{k\in\Z^n}a_k\,U_k$, where $\{a_k\}\in\SS(\Z^n)$ with
the Weyl elements defined by $U_k:=e^{-\frac i2 k.\chi
k}\,u_1^{k_1}\cdots
u_n^{k_n}$, $k\in\Z^n$. Previous relation on the $u's$ reads
\begin{equation}
\label{rel1}
U_{k}U_{q}=e^{-\frac i2 k.\Theta q} \,U_{k+q}, \text{ and }
U_{k}U_{q}=e^{-i k.\Theta q} \,U_{q}U_{k}
\end{equation}
where $\chi$ is
the matrix restriction of $\Theta$ to its upper triangular part.
Thus unitary operators $U_{k}$ satisfy $U_{k}^*=U_{-k}$ and
$[U_{k},U_{l}]=-2i\sin(\frac 12 k.\Th l)\,U_{k+l}$.

Let $\tau$ be the trace on $\A_\Th)$ defined by
$\tau\big( \sum_{k\in\Z^n}a_k\,U_k \big):=a_0$
and $\H_{\tau}$ be the GNS Hilbert space obtained
by completion of $ \A_\Th$
with respect of the norm induced by the scalar product
$\langle a,b\rangle:=\tau(a^*b)$.
On $\H_{\tau}=\set{\sum_{k\in\Z^n}a_k\,U_k \, : \, \{a_{k}\}_{k} \in
l^2(\Z^n) }$, we have the left and right regular
representations of $\A_\Th$ by bounded operators, denoted respectively by $L(.)$ and $R(.)$.

Let also $\delta_\mu$, $\mu\in \set{1,\dots,n}$, be the $n$ (pairwise
commuting)
canonical derivations, defined by
\begin{equation}
\delta_\mu(U_k):=ik_\mu U_k. \label{dUk}
\end{equation}
 $\A_{\Th}$
acts on $\H:=\H_{\tau}\otimes \C^{2^m}$ where $m=\lfloor \tfrac n2 \rfloor$,  the square integrable sections of the trivial spin bundle over $\T^n$: each element of $\A_{\Th}$ is represented on $\H$ as
$L(a)\otimes1_{2^m}$.

The Tomita conjugation $J_{0}(a):=a^*$ satisfies $[J_{0},\delta_{\mu}]=0$ and  we define $J:=J_{0}\otimes C_{0}$ where $C_{0}$ is an operator on $\C^{2^m}$. The Dirac operator is then given by the selfadjoint extension of 
$$
\DD:=-i\,\delta_{\mu}\otimes \gamma^{\mu},
$$
where we use hermitian Dirac matrices $\gamma$. This implies $
C_{0}\ga^{\alpha}=-\eps \ga^\alpha C_{0}$, 
and $\DD\ U_k \otimes e_i = k_\mu U_k \otimes \gamma^\mu e_i$, 
where $\set{e_i}$ is the canonical basis of $\C^{2^m}$. Moreover,
$C_{0}^2=\pm 1_{2^m}$
depending on the parity of $m$. Finally, one introduces the chirality
(which in the
even case is $\chi:=id \otimes (-i)^{m} \gamma^1 \cdots \gamma^{n}$)
and this yields
that $(\A_{\Th},\H,\DD,J,\chi)$ satisfies all axioms of a spectral
triple, see
\cite{Book,Polaris}.

For every unitary $u \in \A$,
$uu^{*}=u^{*}u=U_{0}$, the perturbed Dirac operator $V_{u}\DD V_{u}^*$ by
the unitary $V_{u}:=\big(L(u)\otimes 1_{2^m}\big)J\big(L(u)\otimes
1_{2^m}\big)J^{-1}$, must satisfy condition (\ref{Jcom}). 

This yields the necessity of a symmetrized covariant Dirac operator:
$\DD_{A}:=\DD + A + \epsilon J\,A\,J^{-1}$ 
since $V_{u}\DD V_{u}^{*}=\DD_{L(u)\otimes 1_{2^m}[\DD,L(u^{*})
\otimes 1_{2^m}]}$:
in fact, for $a \in \A_{\Th}$, using $J_{0}L(a)J_{0}^{-1}=R(a^*)$,
we get $$\epsilon J\big(L(a)\otimes
\gamma^{\alpha}\big)J^{-1}=-R(a^*)\otimes \gamma^{\alpha}.$$
This induces some covariance property for the Dirac operator:
one checks that for all $k \in \Z^{n}$,
\begin{align}
\label{puregauge1}
L(U_{k})\otimes 1_{2^m}[\DD,L(U_{k}^{*})\otimes 1_{2^m}]&=1\otimes
(-k_{\mu}\ga^{\mu}),
\end{align}
so with $C_{0}\ga^{\alpha}=-\eps \ga^\alpha C_{0}$, we get $U_{k}[\DD,U_{k}^{*}]+\epsilon
JU_{k}[\DD,U_{k}^{*}]J^{-1}=0$ and
\begin{align}
\label{covariance}
V_{U_{k}} \,\DD \, V_{U_{k}}^{*}=
\DD=\DD_{L(U_{k})\otimes 1_{2^m}[\DD,L(U_{k}^{*})\otimes 1_{2^m}]}.
\end{align}
Moreover, we get the gauge transformation:
\begin{align}
\label{gaugeDirac}
V_{u} \DD_{A} V_{u}^{*}= \DD_{\gamma_{u}(A)}
\end{align}
where the gauged transform one-form of $A$ is
\begin{align}
\label{gaugetransform}
\gamma_{u}(A):=u[\DD,u^{*}]+uAu^{*},
\end{align}
with the shorthand
$L(u)\otimes 1_{2^m} \longrightarrow u$.

As a consequence, the spectral action is gauge invariant: 
$\SS(\DD_{A},\Phi,\Lambda)=\SS(\DD_{\gamma_{u}(A)},\Phi,\Lambda)$.

An arbitrary selfadjoint one-form $A$, can be written as
\begin{equation}
\label{connection}
A = L(-iA_{\alpha})\otimes\gamma^{\alpha},\,\, A_{\alpha}
=-A_{\alpha}^* \in
\A_{\Th},
\end{equation}
thus
\begin{equation}
\label{dirac}
\DD_{A}=-i\,\big(\delta_{\alpha}+L(A_{\alpha})-R(A_{\alpha})\big)
 \otimes \gamma^{\alpha}.
\end{equation}
If $\tilde A_{\alpha}:=L(A_{\alpha})-R(A_{\alpha})$, 
we get
$\DD_{A}^2=-g^{{\alpha}_{1} {\alpha}_{2}}(\delta_{{\alpha}_{1}}+\tilde
A_{{\alpha}_{1}})(\delta_{{\alpha}_{2}}+\tilde
A_{{\alpha}_{2}})\otimes 1_{2^m} - \tfrac 12
\Omega_{{\alpha}_{1} {\alpha}_{2}}\otimes \gamma^{{\alpha}_{1}
{\alpha}_{2}}
$
where $\gamma^{{\alpha}_{1} {\alpha}_{2}}
:=\tfrac 12(\gamma^{{\alpha}_{1}}\gamma^{{\alpha}_{2}}
-\gamma^{{\alpha}_{2}}\gamma^{{\alpha}_{1}})$, 
$\Omega_{{\alpha}_{1} {\alpha}_{2}}
:=[\delta_{{\alpha}_{1}}+
\tilde A_{{\alpha}_{1}},\delta_{{\alpha}_{2}}
+\tilde A_{{\alpha}_{2}}]\,
=L(F_{{\alpha}_{1} {\alpha}_{2}}) - R(F_{{\alpha}_{1} {\alpha}_{2}})$
with
\begin{align}
    \label{Fmunu}
F_{{\alpha}_{1} {\alpha}_{2}}:=\delta_{{\alpha}_{1}}(A_{{\alpha}_{2}})
-\delta_{{\alpha}_{2}}(A_{{\alpha}_{1}})+[A_{{\alpha}_{1}},A_{{\alpha}_{2}}].
\end{align}
In summary,
\begin{align}
\label{D2}
\DD_{A}^2=-\delta^{{\alpha}_{1} {\alpha}_{2}}
\Big(
\delta_{{\alpha}_{1}}+L(A_{{\alpha}_{1}})-R(A_{{\alpha}_{1}})\Big)
\Big(\delta_{{\alpha}_{2}}+L(A_{{\alpha}_{2}})-R(A_{{\alpha}_{2}})\Big)
\otimes 1_{2^m}
\nonumber\\
-\tfrac 12\,\big(L(F_{{\alpha}_{1} {\alpha}_{2}}) - R(F_{{\alpha}_{1}
{\alpha}_{2}})\big)
\otimes \gamma^{{\alpha}_{1} {\alpha}_{2}}.
\end{align}

\subsection{Spectral action}
We first identify the dimension spectrum:
\begin{theorem}
(i) $Sd(\A_\Th,\H,\DD)=\set{n-k : n \in \N_0}$ and all poles are simple.

(ii) $\zeta_D(s)=2^m\sum_{k \in \Z^n \setminus 0} \vert k \vert^{-s}+{\rm dim }\,\Ker \, \DD$. In particular, $\zeta_D(0)=0$.
\end{theorem}
The reader should notice that 

\centerline{HIC SUNT DRACONES}
In fact, {\it $(i)$ is true when $\Th$ is badly approximable.}

\begin{definition}\label{ba}
(i) Let $\delta >0$. A vector $a \in \R^n$ is said to be
$\delta-$badly approximable
if there exists $c >0$ such that $|q . a -m| \geq c \,|q|^{-\delta}$,
$\forall q \in \Z^n \setminus \set{0}$ and $\forall m \in \Z$. \\
We note ${\cal BV}(\delta )$ the set of $\delta-$badly approximable
vectors and ${\cal
BV} :=\cup_{\delta >0} {\cal BV}(\delta)$ the set of badly
approximable vectors.\par
(ii) A matrix $\Th \in {\cal M}_{n}(\R)$ (real $n \times n$ matrices)
will be
said to said badly approximable if there
exists $u \in \Z^n$ such that  ${}^t\Th (u)$ is a badly approximable
vector of $\R^n$.
\end{definition}
\begin{remark} A classical result from Diophantine approximation
asserts
that for all $\delta >n$, the Lebesgue measure of
$\R^n \setminus {\cal BV}(\delta)$ is zero (i.e almost any element of
$\R^n$ is $\delta-$badly approximable.)
\par Let $\Th \in {\cal M}_n(\R)$. If its row
of index $i$ is a badly approximable vector of $\R^n$ (i.e. if $L_i
\in {\cal BV}$), then ${}^t \Th (e_i) \in {\cal BV}$ and thus $\Th$ is a badly
approximable matrix. It
follows that almost any matrix of ${\cal M}_n(\R)\approx \R^{n^2}$ is
badly approximable.
\end{remark}
This difficulty is due to the presence of $J$: one must control the holomorphic behavior of few Hurwitz--Epstein Zeta functions.
 \begin{theorem} Define
 $$
 f_a: s\in \C \rightarrow \sum_{0 \neq k\in \Z^n} \tfrac{P(k)}{\vert \vert k \vert \vert^s}\, e^{i2\pi\, k.a}
 $$
 where $a\in \R^n$, $P\in \C[x_1,\ldots,x_n]$ is a homogeneous polynomial of degree $p$ and  $\vert \vert k\vert\vert^2=\sum_{i=1}^n {k_i}^2$.
 
 (i) When $a\in \Z^n$, $f_a$ has meromorphic extension to the whole complex plane $\C$ and
 
\centerline{ $f_a$ not entire $\Leftrightarrow \,\Res_{s=n+p} \, f_a(s)=\int_{u \in S^{n-1}} P(u) \, dS(u) \neq 0$.}
 
 (ii)  When $a \in \R \setminus \Z^n$, $f_a$ extends holomorphically to the whole complex plane $\C$.
 
  (iii) When $\Th$ is badly approximable, for any integer $q>0$, the function
 $$
 g(s):=\sum_{l \in ({\Z^n)}^q} c(l) \, f_{\Th (\sum_i \epsilon_i\,l_i)}(s), \quad \text{ with  } c(l) \in \SS{(\Z^n)}^q \text{ and } \epsilon_i \in \set{1,0,-1}.
 $$
 extends meromorphically to $\C$ with only one possible pole at $s=n+p$. If this pole exists, it is simple and
 $$
 \Res_{s=n+p} \, g(s)=c \int_{u\in S^{n-1}}P(u)\, dS(u) , \text{ with } c:=\sum_{l \in Z} b(l) \text{ where } Z:=\set{l \in {(\Z^n)}^q : \sum_{i=1}^q \epsilon_i l_i=0}.
 $$
\end{theorem}
Examples: 
\begin{align*}
&\Res_{s=0} \, \sum_{k\in \Z^2} \frac{k_i k_j}{\vert \vert k \vert \vert^{s+4}}=\delta_{i,j} \,\pi \, ,\\
&\Res_{s=0} \, \sum_{k\in \Z^4} \frac{k_i k_j k_l  k_m}{\vert \vert k \vert \vert^{s+6}}=(\delta_{ij}\delta_{lm}+\delta_{il}\delta_{jm}+\delta_{im}\delta_{jl} )\,\tfrac{\pi^2}{12 \, },
\end{align*}
with other similar results, see \cite{EILS} for details. This might be helpful for computations in $\zeta$-regularization, multiplicative anomalies or Casimir effect.

The main result is (see \eqref{Fmunu} for notations)
\begin{theorem}
Assume $\tfrac{1}{2\pi} \Th$ is badly approximable. For any selfadjoint one-form A, the spectral action of the noncommutative torus of dimension n is 

for n=2:
$$\SS(\DD_A,\Lambda,\Phi)= 4\pi\,\Phi_2 \,\Lambda^2+\mathcal{O}(\Lambda^{-2}),$$

for n=4:
$$\SS(\DD_A,\Lambda,\Phi)= 8\pi^2\, \Phi_4 \, \Lambda^4-\tfrac{4\pi^2}{3}\,\Phi(0)\, \tau(F_{\mu \nu} F^{\mu \nu})+\mathcal{O}(\Lambda^{-2}).$$

More generally, $\forall n \geq 1$,
\begin{align}
\label{result}
\SS(\DD_A,\Lambda,\phi)=\sum_{k=0}^n \Phi_{n-k}\, c_{n-k}(A) \, \Lambda^{n-k}+\mathcal{O}(\Lambda^{-1})
\end{align}
with $c_{n-1}(A)=0$, $c_{n-k}(A) = 0$ for $k$ odd (n odd $\Rightarrow c_0(A)=0.)$
\end{theorem}

\begin{conjecture} The ratio \big[constant term of $\SS(\DD_A,\Lambda,\phi)\big]$/\big[constant term of $\SS(\DD+A,\Lambda,\phi)$ for the commutative torus (i.e. $\Th=0$)\big] is independent of $A$. Note, however that, $\DD_A=\DD, \, \forall A=A^*$ in the commutative torus.
\end{conjecture}

\begin{remark}
(i) For general spectral triples with simple dimension spectrum, $\ncint$ can be defined with $\DD$ or $\DD_A$:
$$
\ncint P= \Res_{s=0} \, \Tr \big( P\vert D_A \vert ^{-s} \big), \quad \forall P \in \Psi DO.
$$
(ii) The top term (cosmological term) is covariance-invariant: $\ncint \vert D_A \vert ^{-n}=\ncint \vert D \vert ^{-n}$.

(iii) There is no tadpole term in \eqref{constant1} for the noncommutative torus: $\ncint \wt{A} D^{-1}=0$.

(iv) A Diophantine condition was characterized by Connes in \cite[Prop. 49]{NCDG} for $n=2$ in terms of Hochschild cohomology
$H(\A_{\Th},{\A_{\Th}}^*)$.
\end{remark}

\subsection{Beyond Diophantine condition}
It is interesting to overcome the Diophantine condition and to consider for instance the case $n=2$, where
$\Th=\th\genfrac{(}{)}{0pt}{1}{\,0 \quad1}{-1\,\,\,\, 0 }$
with $\th \in \R$. 

Let $f: [1,\,\infty[ \rightarrow ]0,\,\infty[$ be a continuous function such that  $x^2\,f(x)$ is non-increasing.
$$\mathcal{K}(f):=\set{\th \in \R \, : \, \vert \th -\tfrac{p}{q} \vert < f(q), \text{ for infinitely many rational numbers } \tfrac{p}{q} }$$
Such $\th$ are termed $f$-approximable.
 Note that this is not valid for all rational $\frac pq$ since $(\th q)_{q\geq 1}$ are dense in $[0,\,1]$ when $\th \notin \Q$.

\begin{lemma}
 Jarnik (1953): For each $f$, there exists an uncountable set $\mathcal{K}(f)$ of real numbers $\tfrac{1}{2\pi} \th$, $f$-approximable but not $c\, f$-approximable for any $0<c<1$.
 \end{lemma}

$\mathcal{K}(f)$ has zero Lebesgue measure if $\sum_{q=1}^\infty q\,f(q)$ converges and full Lebesgue measure otherwise.
\vskip0.2cm
Consequences: Tuning $f$, $\exists \, a,\, b\in \A$ such that the 
correction term  $$\Tr \big( L(a)\,R(b)\, e^{-t\DD^2} \big)-\big(\text{same term when $\Th$ badly approximable}\big)$$
is not exponentially small, is not $\mathcal{O}(\frac 1t)$ like for $\tfrac{1}{2\pi} \th \in \Q$, but can be of arbitrary order!

This could have consequences on computations in field theories over noncommutative tori, see \cite{GIVas} for details.

Naturally, all these complications are related to the compactness of the torus, so it is natural to consider the

\section{Extension to noncompact manifolds}

Let $M$ be a $n$-dimensional non compact connected complete Riemannian spin$^c$ manifold with bounded curvature and control of its heat kernel $K_t(.,\,.)$ like
\begin{align*}
& \sup_{p\in M} \int_0^\infty t^k\, e^{-t} K_t(p,p) \, dt < \infty, \quad \forall k>\tfrac n2-1\\
& \sup_{p\in M} \int_m^\infty \tfrac{e^{-t}}{\sqrt{t}} \, K_t(p,p) \, dt < c \, m^{-(n-1)/2}, \quad \forall m \in [0,\, 1]
\end{align*}
This hypotheses are valid in the following cases (see \cite{GIV}):

- $M$ has Ricci curvature bounded from below.

- $M$ has a positive injectivity radius and control of isoperimetric constants of balls of a given radius.

- $M$ has a bounded geometry.

Moreover, $M$ is given a smooth isometric proper action $\alpha$ of $\R^l$.

Example: The Moyal planes where  $M=\R^{2n}$ and $\alpha=$ translation. A Moyal multiplication can be defined for any $n \times n$-skew-symmetric matrix $\Th$ by
$$
f \star_{\hbar} g(x):=\int_{\R^{2n} \times \R^{2n}} f(x-\Th u)\,g(x-v) \, 
e^{-i\,u\cdot v}\,du\,dv.
$$

\begin{theorem} \cite{GIV} 
When $f\in C_c^\infty(M)$, 
$\ncint L_f \vert \DD\vert^{-d}=\ncint M_f \vert \DD\vert^{-d}=c\int_M f.$
\end{theorem}
Remarks:

- When $M$ is compact, $\alpha$ being proper must be periodic.

- When $M$ is non compact and as above, $\DD$ has a continuous spectrum but for any $f\in C_c^\infty(M)$, and any $p\geq 2$, 
$L_f(1+\vert \DD \vert)^{-2k} \in \L^p(\H), \, \forall k>\tfrac{n}{2p}$. 
This particularly fit our purpose since for noncompact spectral triple, we must have by hypothesis 
\begin{align}
\label{rescompact}
a\, (\DD-\lambda)^{-1} \text{ is compact } \forall a \in \A.
\end{align}
This is indeed the case for Moyal planes:
\begin{theorem}
\cite{GGIV} 
Moyal planes are spectral triples.
\end{theorem}

To compute a spectral action in the noncompact case, one can use \eqref{rescompact} for instance with some spatial localization:  Given $\rho \in \A$, the action is defined by 
\begin{align}
\label{defnc}
\SS(\DD,\Lambda,\Phi,\rho):=\Tr \big(\rho \, \Phi(\DD/\Lambda) \big).
\end{align}
In this setting,

\begin{theorem} 
\cite{GI2002} 
$\SS(\DD,\Lambda,\Phi,\rho)$ has same coefficients as in \eqref{result} with replacement $c_k(A)\leftrightarrow \int_{\R^{n/2}}\rho(x)\, c_k(A)(x) \, dx$.
\end{theorem}

\begin{remark}
Definition \eqref{defnc} is not satisfactory since there are too many choices. This question has been investigated in \cite{CC2} where a dilaton field $\phi$ is used and the squared Dirac operator $\DD^2$ is replaced by $e^{-\phi}\, \DD^2 \, e^{-\phi},\, \phi=\phi^* \in Z(\A)$. Thus in spectral action, the counting of eigenvalues $N(\Lambda):=\dim \set{\DD^2 \leq \Lambda^2}$ is replaced by $N(\rho):=\dim \set{\DD^2 \leq \rho^2}$ where $\rho=e^\phi$.

A general framework for a good definition of spectral action in the noncompact setting is still lacking.
\end{remark}

\ack
It is a pleasure to thank Driss Essouabri, Victor Gayral, Cyril Levy and Andrzej Sitarz and Joe V\'arilly for collaborations on which this review is based.

\end{document}